\documentclass[two-column,10pt]{article}
\usepackage[a4paper,top=2cm,bottom=3cm,left=3cm,right=3cm]{geometry}
\usepackage{titlesec}
\usepackage{amsmath,amsfonts,amssymb}
\usepackage{graphicx}
\usepackage{hyperref}
\usepackage{titlesec}
\usepackage{fancyhdr}
\usepackage{authblk}
\usepackage{abstract}
\usepackage{float}
\usepackage{titlesec}
\usepackage{subcaption}
\usepackage{multirow}
\usepackage{array}

\usepackage[numbers,compress]{natbib} 
\titleformat{\section}{\large\bfseries}{\thesection}{1em}{}
\titleformat{\subsection}{\normalsize\bfseries}{\thesubsection}{1em}{}

\title{\vspace{0.3cm}\textbf{Randomness from Radiation: Evaluation and Analysis of Radiation-Based Random Number Generators}}

\author[1]{Roohi Zafar}
\author[2*]{Muhammad Kamran}
\author[3]{Tahir Malik}
\author[1]{Kashish Karera}
\author[1]{Humayon Tariq}
\author[1]{Ghulam Mustafa}
\author[2]{Muhammad Mubashir Khan}

\affil[1]{Department of Physics, NED University of Engineering \& Technology, University Road, Karachi-75270, Pakistan}

\affil[2]{Department of Computer Science \& Information Technology, NED University of Engineering \& Technology, University Road, Karachi-75270, Pakistan}

\affil[3]{Department of Telecommunication Engineering, NED University of Engineering \& Technology, University Road, Karachi-75270, Pakistan}

\affil[ ]{\vspace{0.5cm}}

\affil[*]{\textit{Muhammad Kamran. E-mail(s): kamran@cloud.neduet.edu.pk}}
\affil[ ]{\textit{Contributing authors: roohizj@cloud.neduet.edu.pk; tmalik@cloud.neduet.edu.pk; 
gmkhan@neduet.edu.pk;
mmkhan@cloud.neduet.edu.pk; kashishkarera@gmail.com}}
\affil[ ]{\textit{†These authors contributed equally to this work.}}

\begin{document}

\twocolumn[
\begin{@twocolumnfalse}
    \noindent\maketitle
\end{@twocolumnfalse}

\begin{onecolabstract}
Random numbers are central to various applications such as secure communications, quantum key distribution theory (QKD), statistics, and other tasks. One of today's most popular generators is quantum random numbers (QRNGs). The inherent randomness and true unpredictability in quantum mechanics allowed us to construct QRNGs that are more accurate and useful than traditional random number generators. Based on different quantum mechanical principles, several QRNGs have already been designed. The primary focus of this paper is the generation and analysis of quantum random numbers based on radioactive decay. In the experimental set, two beta-active radioactive sources, cobalt-60 (Co60) and Strontium-90 (Sr 90) , and an ST-360 counter with a Geiger-Muller (GM) tube are used to record the counts. The recorded data was then self-tested by entropy and frequency measurement. Moreover, popular testing technique, the National Institute of Science and Technology (NIST) randomness testing is used , to ensure that the guaranteed randomness meets security standards. The research provides the impact of the nature of the radioactive source, the distance between the counter and sources, and the recording time of the counts on generating quantum random numbers of radioactive QRNGs.
\end{onecolabstract}
\vspace{1em}
]
\section{Introduction}
\begin{figure}
    \centering
    \includegraphics[width=1\linewidth]{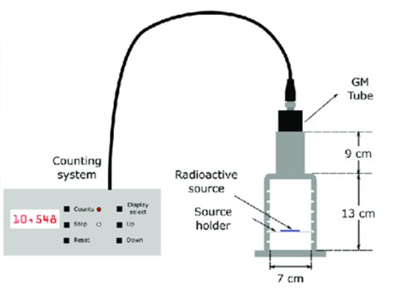}
    \caption{Schematic diagram of experimental setup}
    \label{fig:enter-label-1}
\end{figure}

Advancements in quantum mechanics have introduced new protocols that link telecommunication, information technology, physics, and computer science. Advancements in technology, such as quantum key distribution protocols (QKD) \cite{bennett1984update, ekert1991quantum, khan2017detailed, nurhadi2018quantum} and practical algorithms \cite{childs2010quantum, ekert1996quantum}, highlight the vital impact that quantum physics can have on unconditionally secure communication, cryptography, and computation. Quantum random number generators (QRNG) \cite{manucom2019analysis} are another essential technology in this realm. Random number generators (RNGs) \cite{l2014random} can be characterized into two main categories: pseudo-random number generators (PRNGs), also known as software RNGs, and true random number generators (TRNGs) \cite{stipvcevic2014true}. Many researchers have sought to validate randomness based solely on the analysis of observed random sequences \cite{kolmogorov1998tables, marsaglia1996diehard, kim2004corrections}. However, generating truly random numbers has proven difficult in classical computing. Based on generation phenomena, TRNGs can be further divided into two categories: physical random number generators and quantum random number generators. In physical random number generators, random numbers are generated by the measurement of classical system parameters with disordered behavior, but at the fundamental level, numbers are specific, where as, in quantum random number generators (QRNGs), the random numbers are generated based on the inherent uncertainty in a quantum system \cite{zettili2009quantum}. QRNG devices are used in various fields and are generally more straightforward than other quantum technologies for practical applications \cite{wilber2013entropy, bell1964einstein, owens2008entangled, bouda2012weak, sharma2020quantum, shannon1949communication}. 
\newline 
Several QRNGs exist based on different quantum mechanical principles, and each of the different QRNG designs leverage and exploit different quantum systems\cite{wayne2009photon, applegate2015efficient, mannalatha2023comprehensive, shen2010practical, liu2018entanglement}. The first ever Quantum Random Number Generator(QRNG) was based on radioactive decay, the simplest method of generating quantum random numbers \cite{herrero2017quantum}. Radioactive decay is the change of an atomic nucleus from an unstable state to a more stable state through the release of particles or energy\cite{groch1998radioactive}. This process follows the law of quantum mechanics, and the lifetime of an atom or any other radioactive substance cannot be predicted and can only happen statistically\cite{gurney1929quantum}. This unpredictability is used by QRNGs, which count particles or photons produced due to radioactive decay; timing and counting are used to generate random numbers  \cite{lavine2000decay}. Almost all the quantum number generators that depend on the decay feature have sensitive detectors, allowing them to determine decay events at the individual event level accurately. These usually are radiation counters like Geiger-Muller tubes and ionization chambers that detect and convert the radiation emitted into electric signals. These signals are then subjected to the computation software, where they are recorded and converted into bits of random numbers. In this research paper, different experiments are designed to analyze the effect of different parameters, like the half-life of the source, different recording times, etc., that might affect the randomness of QRNGs. The experimental setup consists of radioactive sources, an ST-360 radiation counter with a GM Tube; a gas-filled tube that gets ionized upon interaction with radiation, producing pairs of positive ions and free electrons that are further accelerated to create an avalanche effect, a power supply transformer, and a stand. The schematic diagram is shown in Figure~\ref{fig:enter-label-1}. The first experiment is designed to calculate the operating voltage of the GM counter using the Geiger plateau graph. To see the effect of different parameters that might out turn the randomness of the QRNGs, the counts are recorded for different sources, preset times, and distances of sources from the GM counter to see the effect of these parameters on the randomization of random numbers. We conduct rigorous testing, including statistical randomness tests and NIST SP 800-22 tests \cite{bassham2010sp800}, to ensure the randomness meets strict security standards.

\section{Methodology}
\begin{figure}[H]
    \centering    \includegraphics[width=1\linewidth]{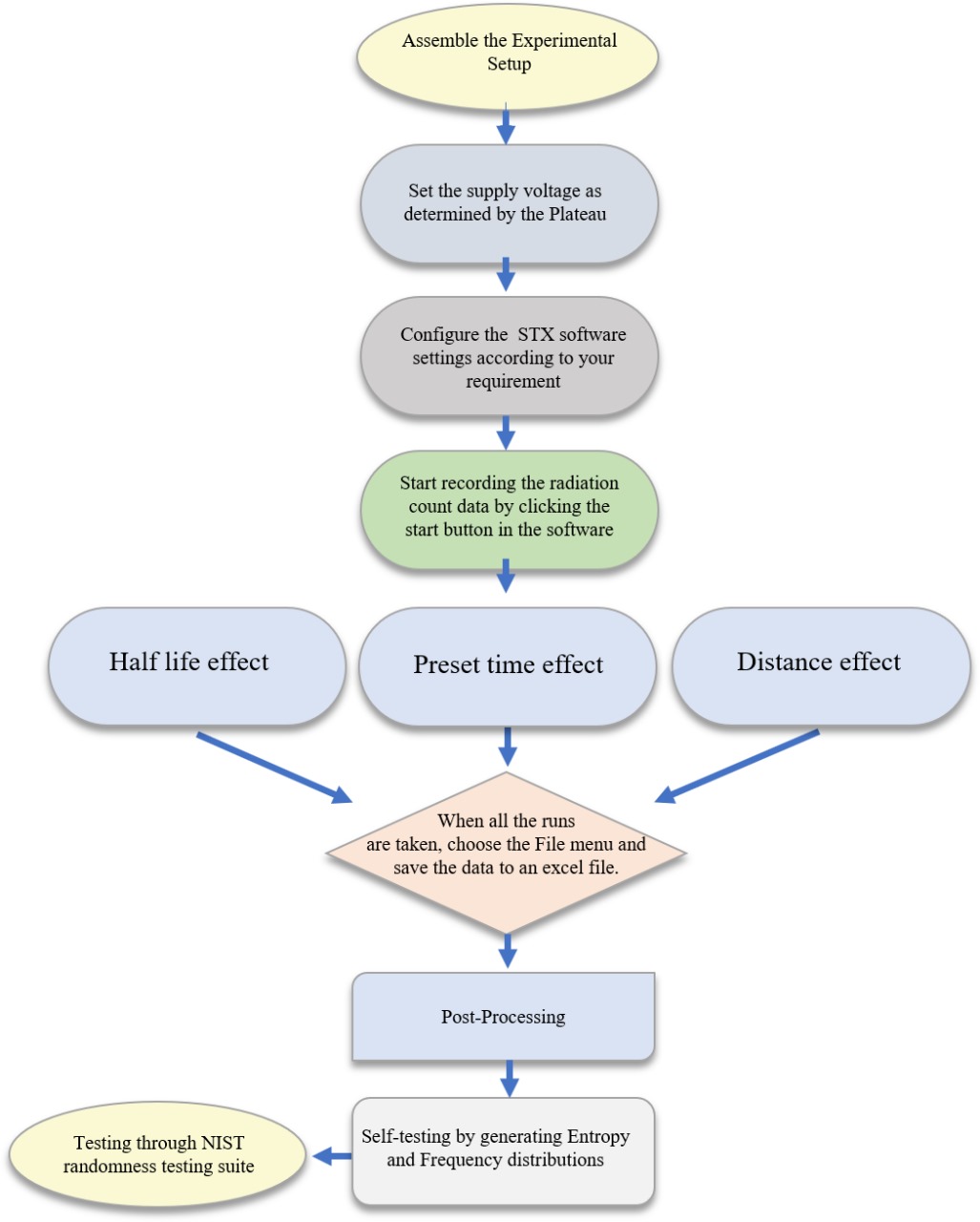}
    \caption{Overview of the Methodology}
    \label{fig:enter-label-2}
\end{figure}

The research methodology uses different experimental setups, post-processing, and testing techniques. Figure~\ref{fig:enter-label-2}  shows the flow chart of the method used to analyze the experiment's results regarding the randomness of the data.
\subsection{Experimental Setup}
\begin{figure}
    \centering    \includegraphics[width=1\linewidth]{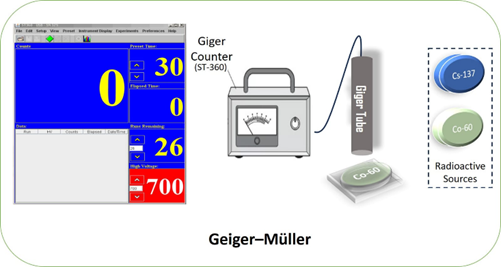}
    \caption{Schematic Diagram of ST-360 setup }
    \label{fig:enter-label-3}
\end{figure}

Initially, the power supply is turned on to turn the GM counter on, and the GM tube is carefully placed on the top of the shelf stand without touching its window. Then, the GM tube and ST-360 are connected using a Bayonet NeillConcelman (BNC) cable, and a USB cable is attached to the ST-360 and computer, as shown in the schematic diagram of the experimental setup with interface Figure~\ref{fig:enter-label-3}.

\begin{figure}[H]
    \centering
    \includegraphics[width=0.85\linewidth]{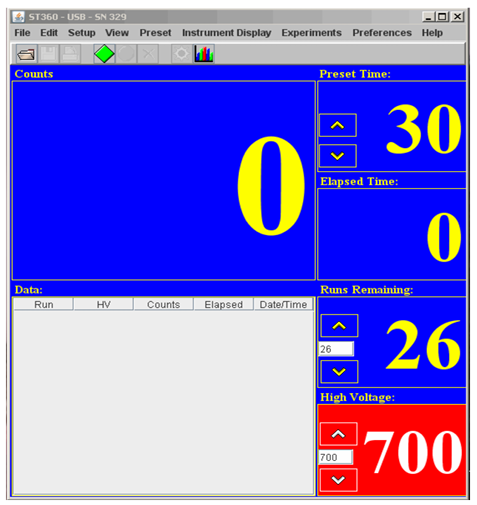}
    \caption{\vspace{0.2cm}The window of ST-360 software\\\textbf{Note:}
    This is just an example window, not the actual settings. The interface shows that the tube counts radiations for 30 seconds, the voltage is set to 700 and the total number of runs are set to 26.}
    \label{fig:enter-label-4}
\end{figure}

Every GM counter operates differently because it is constructed uniquely. Therefore, the initial step is to determine the operating voltage of ST-360. To determine the operating voltage of ST-360, the voltage across the tube is increased by 20 volts, and the count rate is recorded every half a minute (30 seconds).
After setting up the optimal power supply voltage, we move to the software settings, where the counts are imported from ST-360. The screen appears as shown in Figure~\ref{fig:enter-label-4}.
The interface represents number of detected counts, remaining runs, the window of preset time which is when the gas in the GM tube is ionized by radiation events generating pulses that the GM counter counts. After the preset time, the GM counter stops counting the radiation events. The elapsed time window represents how much preset time has passed, and the remaining runs show how many runs are left (Note that the number of runs is pre-defined). Then, the sources are placed beneath the shelf stand at some distance, and the counter is turned on by clicking the green button in the window. Applied Voltage appears on the bottom right. 
\newline
After experimenting with different comparison parameters, half-life of two different sources, preset time, and distances that might affect the randomness of the data, the recorded data is saved in an Excel sheet, or a text file can also be generated, which is used for further post-processing and testing.
\subsection{Post-processing}
\begin{figure}[H]
    \centering
    \includegraphics[width=1\linewidth]{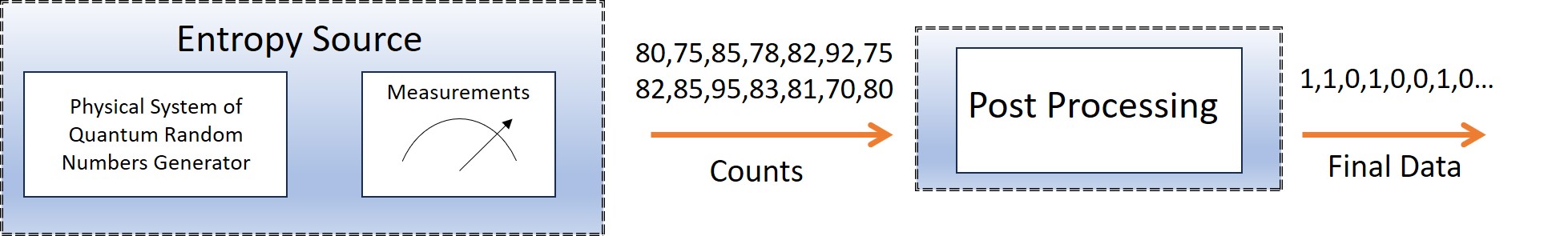}
    \caption{Schematic Diagram of Post-Processing}
    \label{fig:enter-label-5}
\end{figure}

The post-processing technique involves generating random bits by taking the mean of all the counts and implying that if a certain count is greater than the mean, a value of 1 is assigned; otherwise, 0 occurs. The resulting bits are combined into a string by using a fairly simple excel function and subjected towards testing. Figure~\ref{fig:enter-label-5} summarizes the whole post-processing technique while the conditions for post-processing is shown in Table~\ref{tab:bit_condition}.

\vspace{0.5cm}\begin{table}[h!]
\centering
\begin{tabular}{|c|c|c|}
\hline
\textbf{Count ($x_i$)} & \textbf{Mean ($\mu$)} & \textbf{Output} \\ \hline
$x_i > \mu$ & $\mu$ & 1 \\ \hline
$x_i \leq \mu$ & $\mu$ & 0 \\ \hline
\end{tabular}
\caption{Post-Processing Condition for Bit Generation}
\label{tab:bit_condition}
\end{table}

\subsection{Testing}
Initially, a set of 5000 counts of Co-60 and Sr-90 was evaluated by self-testing and, then NIST randomness testing suite SP 800-22 was implied.
\subsubsection{\textit{Self-Testing}}
Primarily, the balance between a number of 0s and 1s was checked by frequency or mono-bit testing technique to ensure a higher chance of randomness. The test was performed by using Python programming language code. To further ensure the randomness quality, entropy testing was performed on both the data by using the Shannon Entropy formula. The formula uses a sliding window approach to calculate the entropy. Window size refers to the data segments for evaluation from a large set of data (5000). Usually, the window size is set to at least 10\% of the size of the data i.e.500. Since entropy is a measure of disorder, randomness, and unpredictability; therefore, the higher the entropy, the higher the randomness. The ideal result would be a straight line at the value of 1; however, the values are expected to fluctuate.
The Shannon entropy \( H(X) \) is defined as:

\[
H(X) = -\sum_{i=1}^{n} p_i \log_2 p_i \tag{\textbf{1}}
\]

where:
 \begin{description}
    \item \( H(X) \) represents the entropy of the random variable \( X \).
    
    \item \( \sum_{i=1}^{n} \) represents summation over all possible outcomes \( i \) from 1 to \( n \).
    
    \item \( p_i \) is the probability of the occurrence of the \( i \)-th unique value within a sliding window (or data set).
    
    \item \( \log_2 \) is the logarithm to the base 2, which gives the entropy in bits. 
  \end{description}

\subsubsection{\textit{NIST Testing}}
Since more than self-testing is needed to judge the quality of randomness, the results cannot be concluded solely on the basis of self-testing. Therefore, the NIST randomness testing technique is applied. NIST's randomness testing suite contains 16 statistical tests that are recognized worldwide to ensure the randomness quality of the data further. A \textit{P}-value (Probability of measurement) of 0.01 is used to decide whether a test is random or not.  
\section{Results and Discussion}
\subsection{Experimental Measurements}
\begin{figure}[H]
    \centering
    \includegraphics[width=1\linewidth]{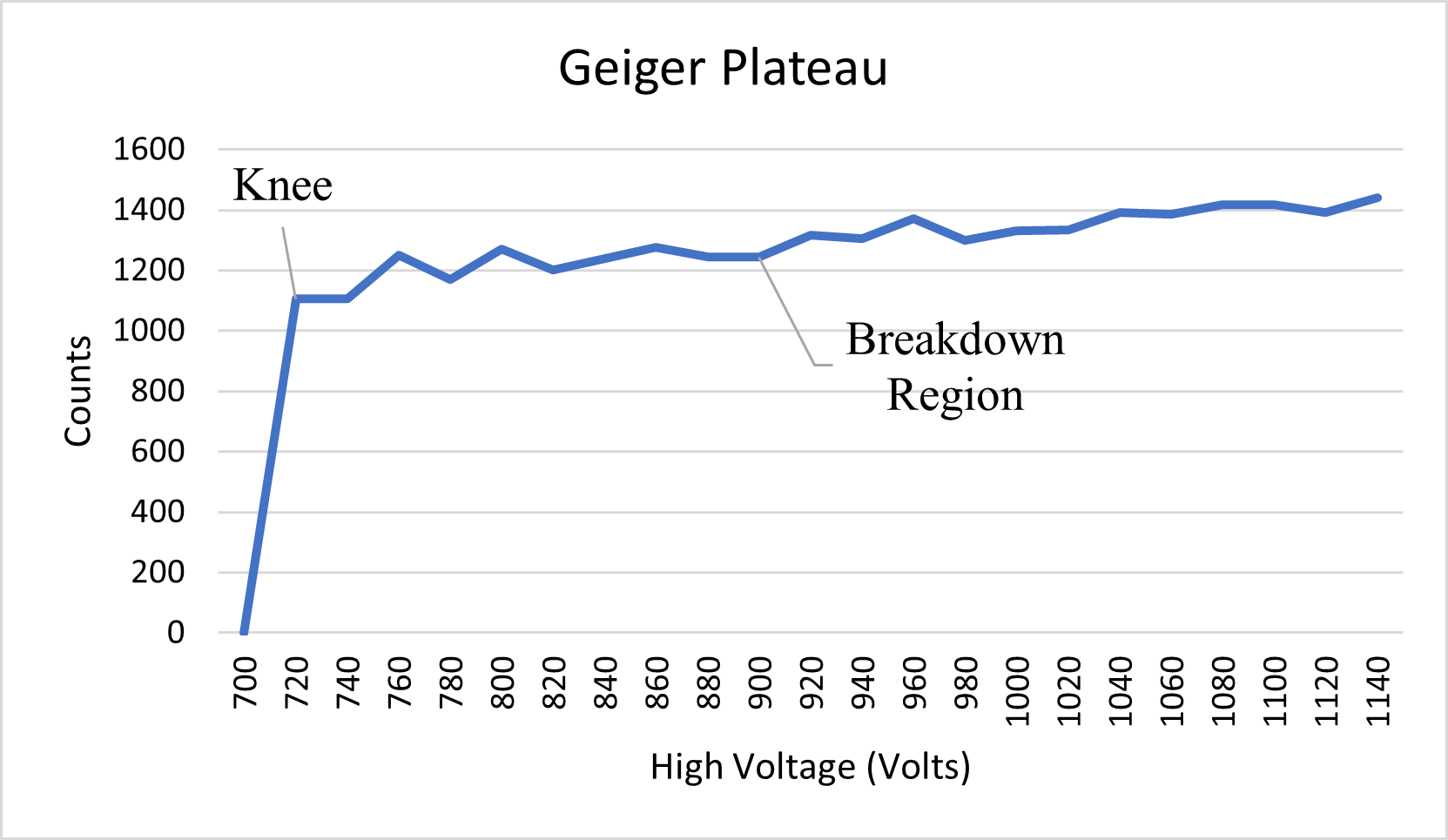}
    \caption{The figure shows the plateau region in between the Knee and Breakdown region}
    \label{fig:Table_2}
\end{figure}

To determine the operating voltage of ST-360, a graph between voltage and count rate is plotted to calculate the GM tube's operating voltage as indicated by Table~\ref{tab:my_label} and plotted in Figure~\ref{fig:Table_2}.
\newline
Initially, the count rate is 0. When the tube reaches a point where avalanche begins it starts counting with a sudden increase in voltage, until it reaches “Knee” or threshold voltage where it starts becoming almost constant despite further increase in the voltage. The best optimal voltage lies in the middle of the plateau region on the voltage curve of ST-360, where the count rate remains relatively constant despite the increased applied voltage, ensuring accurate and stable radiation measurements.

\begin{table}[H]
    \centering
    \begin{tabular}{|c|c|}
    \hline
        \textbf{Parameters} &{Values} \\ 
        \hline
         \textbf{Preset Time} & 30 \\
         \hline
         \textbf{High Voltage} & 700 \\
         \hline
         \textbf{Step Voltage} & 20 \\
         \hline
    \end{tabular}

    \vspace{1cm} 

    \begin{tabular}{|c|c|}
        \hline
        \textbf{High Voltage} & \textbf{Counts} \\
        \hline
        700 & 0\\
        \hline
        720 & 1106\\
        \hline
        740 & 1105\\
        \hline
        760 & 1251\\
        \hline
        780 & 1171\\
        \hline
        800 & 1270\\
        \hline
        820 & 1200\\
        \hline
        840 & 1238\\
        \hline
        860 & 1277\\
        \hline
        880 & 1245\\
        \hline
        900 & 1244\\
        \hline
        920 & 1316\\
        \hline
        940 & 1306\\
        \hline
        960 & 1370\\
        \hline
        980 & 1300\\
        \hline
        1000 & 1330\\
        \hline
        1020 & 1335\\
        \hline
        1040 & 1392\\
        \hline
        1060 & 1386\\
        \hline
        1080 & 1418\\
        \hline
        1100 & 1418\\
        \hline
        1120 & 1392\\
        \hline
        1140 & 1442\\
        \hline
    \end{tabular}
    \caption{Count rate of Co-60 with step voltage of 20}
    \label{tab:my_label}
\end{table}
\vspace{-0.1cm}

\subsection{Effect of Half-life on Randomness}
\begin{figure}[H]
    \centering
    \includegraphics[width=1\linewidth]{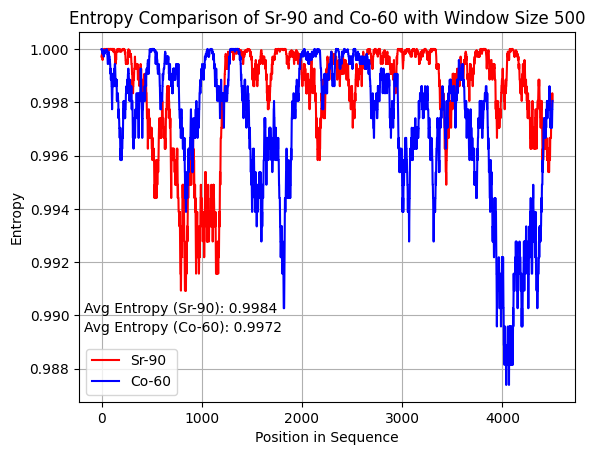}
    \caption{Entropy testing results comparison for the two radioactive sources}
    \label{fig:enter-label-7}
\end{figure}

\begin{figure}[H]
    \centering
    \begin{subfigure}{0.45\textwidth}
        \centering
        \includegraphics[width=\linewidth]{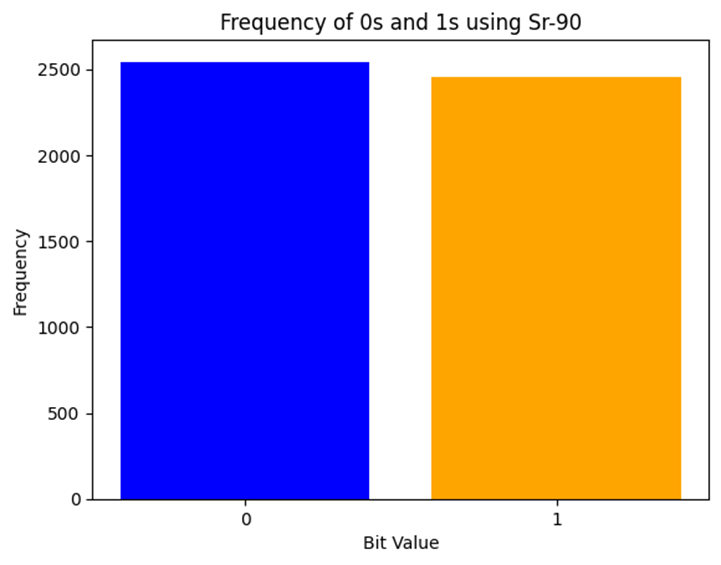}
    \end{subfigure}
    \hfill
    \begin{subfigure}{0.45\textwidth}
        \centering
        \includegraphics[width=\linewidth]{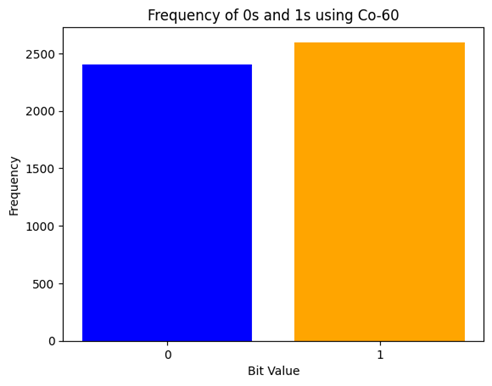}
    \end{subfigure}
    \caption{Comparison between the frequency plotting of the two sources}
    \label{fig:enter-label-8}
\end{figure}

To measure the effect of the source's nature, two sources having different half-lives, i.e., Cobalt-60 (Co-60) having a half-life of 5.3 years and Strontium-90 (Sr-90) with a half-life of 29 years were used in this experiment, and both the results were compared, for a fixed distance and fixed preset time; almost 5000 counts for a preset time of 1 second and a distance of 2cm is recorded for two sources. The recorded data is then passed through the post-processing method and tested using the method reported in the methodology. The results of both elements' entropy testing and frequency testing are given in the Figure~\ref{fig:enter-label-7} and Figure~\ref{fig:enter-label-8}.

Entropy measurement graphs for post-processing data were generated and compared to ensure the quality and unpredictability of random numbers generated by two sources as shown in Figure~\ref{fig:enter-label-7}. In this graph, the tests utilize a sliding window size of 500 (in our case) to calculate the entropy across the various segments of the series. Peaks in the graph highlight the point where data is maximum random. The segments with persistent dips show weak points in the data, indicating potential weakness in randomness. The average entropy of the two sources is written on the graph as well as on the table. From the post-processed entropy plotting of experimental data, it is observed that the data recorded from the two sources remain consistently high randomness. Still, the average entropy of Sr-90, an element having a larger half-life, shows more randomness than the Co-60.  This result is further verified by frequency testing as shown in Figure~\ref{fig:enter-label-8}.

Figure~\ref{fig:enter-label-8} further verified that the balance between 0 and 1 for Sr-90 is greater than Co-60, confirming that the randomness generated by sources with a longer half-life is more suitable for QRNG. Finally, the NIST testing technique is used; 16 statistical tests check the recorded data's randomness. The testing results are reported in the table.
\begin{table}[H]
    \centering
    \scriptsize  
    \begin{tabular}{|c|c|c|}
        \hline
        & \textbf{Cobalt} & \textbf{Strontium} \\
        \hline
        \textbf{Entropy} & 99.72\% & 99.84\% \\
        \hline
        \multirow{2}{*}{\textbf{Frequency}} & 0s = 48\% & 0s = 51\% \\
        & 1s = 52\% & 1s = 49\% \\
        \hline
        \textbf{NIST Randomness} & 68.7\% & 87.5\% \\
        \hline
    \end{tabular}
    \caption{Comparison of Cobalt and Strontium}
    \label{my_table_3}
\end{table}

Table~\ref{my_table_3} summarizes the results that satisfy the expected outcomes. Observe that sources with a longer half-life have more randomness.
\subsection{Effect of Preset time on Randomness}
\begin{figure}[H]
    \centering
    \includegraphics[width=1\linewidth]{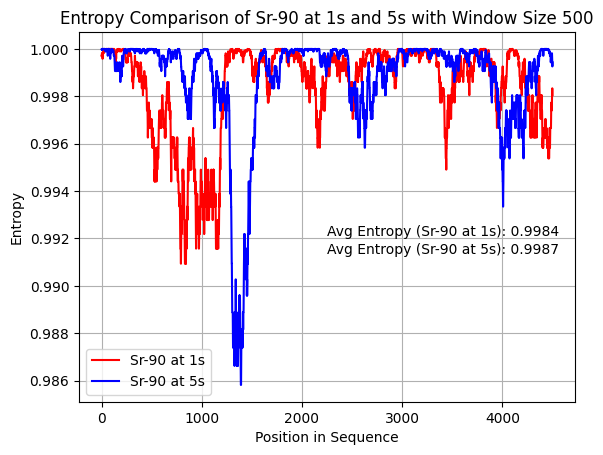}
    \caption{Comparison of Entropy at preset time 1s and 5s using Sr-90}
    \label{fig:enter-label-9}
\end{figure}

\begin{figure}
    \centering
    \begin{subfigure}{0.45\textwidth}
        \centering        \includegraphics[width=\linewidth]{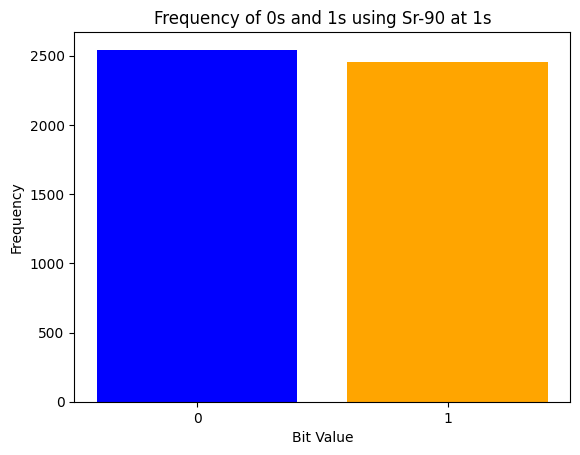}
    \end{subfigure}
    \hfill
    \begin{subfigure}{0.45\textwidth}
        \centering        \includegraphics[width=\linewidth]{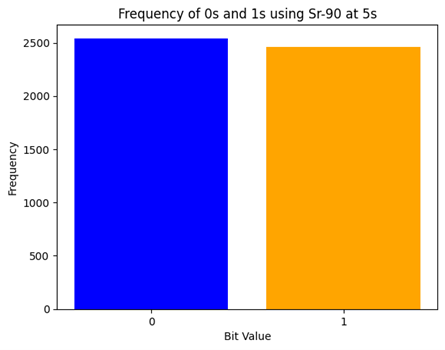}
    \end{subfigure}
    \caption{Comparison between the frequency plotting at the two Preset Times}
    \label{fig:enter-label-10}
\end{figure}

As mentioned above, the preset time refers to the duration for which the tube measures and records the emitted radiation counts. Once the preset time has elapsed, the tube stops counting the radiations, but it does not mean that the radiation emissions have stopped; hence, the preset time should not affect the randomness quality; it should only impact the number of counts. The higher the preset time, the higher the number of counts. Sr-90 was used at two different preset times of 1 second and 5 seconds under similar laboratory conditions, and the 5000 counts were taken each time. The recorded data was plotted on the entropy as shown in Figure~\ref{fig:enter-label-9}.
The observations show that the entropy isn't affected much by present time as both the averages are almost equal. To further verify this, frequency testing can be observed as illustrated in Figure~\ref{fig:enter-label-10}.
Figure`\ref{fig:enter-label-10} doesn't either show much difference. To ensure that the self-testing results are accurate, we move to the NIST testing randomness software.
\begin{table}[H]
    \centering
    \scriptsize  
    \begin{tabular}{|c|c|c|}
        \hline
        & \textbf{Sr-90 at 1s} & \textbf{Sr-90 at 5s} \\
        \hline
        \textbf{Entropy} & 99.84\% & 99.87\% \\
        \hline
        \multirow{2}{*}{\textbf{Frequency}} & 0s = 51\% & 0s = 51\% \\
        & 1s = 49\% & 1s = 49\% \\
        \hline
        \textbf{NIST Randomness} & 87.5\% & 87.5\% \\
        \hline
    \end{tabular}
    \caption{Comparison of Sr-90 at 1s and 5s}
    \label{tab:comparison}
\end{table}

NIST testing also does not show any difference. Hence, the results are satisfied with the expected outcomes.
observes that sources with a higher half-life haveobserves that sources with a higher half-life have
\subsection{Effect of distance of GM tube from the source on Randomness}
\begin{figure}[H]
    \centering    \includegraphics[width=1.1\linewidth]{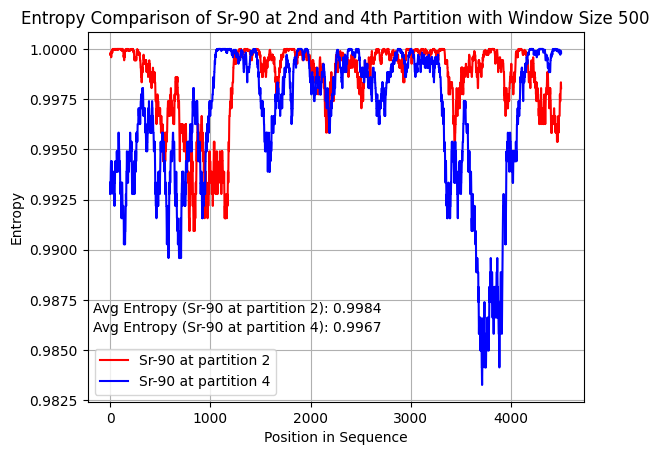}
    \caption{Entropy testing result comparison for the two distances from the GM tube}
    \label{fig:enter-label-11}
\end{figure}

\begin{figure}
    \centering
    \begin{subfigure}{0.45\textwidth}
        \centering
        \includegraphics[width=\linewidth]{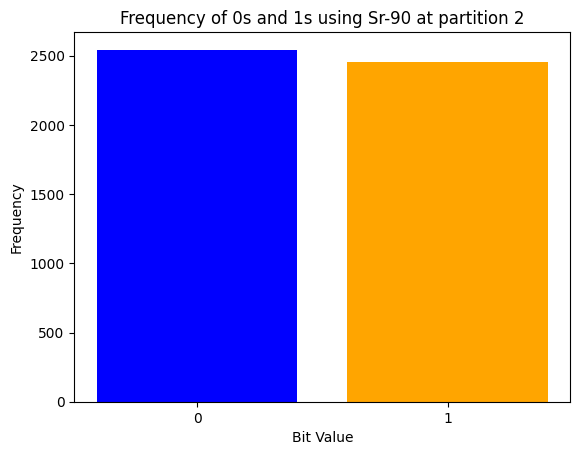}
    \end{subfigure}
    \hfill
    \begin{subfigure}{0.45\textwidth}
        \centering
        \includegraphics[width=\linewidth]{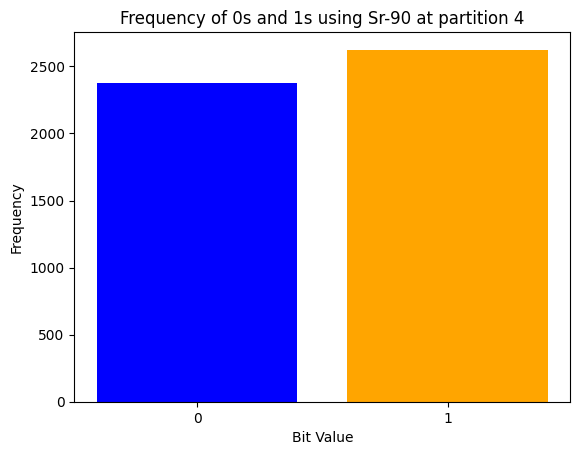}
    \end{subfigure}
    \caption{Comparison between the frequency plotting at the two Distancess}
    \label{fig:enter-label-12}
\end{figure}

The shelf where the source is placed contains partitions at different distances. The total length of the partitions is 13 cm. Sr-90 was first placed on the second partition, and the same experiment was performed on the fourth partition. Due to the larger distance between the GM tube and the source, the electron-electron repulsion(since Sr-90 is a beta-emitting source and beta particles are electrons) will have more space and time to scatter before reaching the tube; therefore, fewer radiations reach the tube, leading to a low detection rate. Moreover, the background noise becomes prominent, resulting in a high noise ratio. Consequently, the quality of the randomness decreases. The preset time was set to 1 second, and 5000 counts were recorded. Entropy and frequency tests were performed as illustrated in Figure~\ref{fig:enter-label-11} and Figure~\ref{fig:enter-label-12} .
Entropy testing shows a significant reduction in the randomness results due to the increment in distance.
\newpage
A noticeable difference can also be illustrated in the frequency testing.
\begin{table}[H]
    \centering
    \scriptsize  
    \begin{tabular}{|c|c|c|}
        \hline
        & \textbf{Sr-90 at 2} & \textbf{Sr-90 at 4} \\
        \hline
        \textbf{Entropy} & 99.84\% & 99.67\% \\
        \hline
        \multirow{2}{*}{\textbf{Frequency}} & 0s = 51\% & 0s = 48\% \\
        & 1s = 49\% & 1s = 52\% \\
        \hline
        \textbf{NIST Randomness} & 87.5\% & 68.75\% \\
        \hline
    \end{tabular}
    \caption{Comparison of Sr-90 at 2\textsuperscript{nd} and 4\textsuperscript{th} partition}
    \label{my_table_5}
\end{table}

In general, Table~\ref{my_table_5} briefly presents the impact of the distance of the radioactive source from the GM tube on the randomness that meets expectations. 
\section{Conclusion}
Ultimately, all the results are satisfactory and show the high-quality randomness used in the QRNG applications. However, the bit-generate rate is quite low and needs to be raised. Due to the half-life constraint, one must use the source having a higher enough half-life to produce good quality randomness at lower preset time for faster bit generation rate. The distance between the source and the GM tube should be kept as low as possible to avoid scattering due to the repulsive nature of beta particles. Lastly, safety measures should not be neglected while working with radioactive sources. 

\section{Acknowledgments}
The authors like to thank NED University of Engineering \& Technology, Department of Physics (Physics Lab-1), for providing the research facilities. Special thanks to Mr. Muhammad Badar Alam for his invaluable assistance in the collection of radioactive particle data.  

\section{Funding}
No funding is acquired for the research work.

\section{Conflict of Interest}
All authors contributed equally in the research work.

\bibliography{main}

\end{document}